\documentclass[fleqn,twoside]{article}
\usepackage[headings]{espcrc2}

\readRCS
$Id: espcrc2.tex,v 1.2 2004/02/24 11:22:11 spepping Exp $
\usepackage{graphicx}
\usepackage[figuresright]{rotating}

\usepackage{axodraw}

\newcommand{\AmS}{{\protect\the\textfont2
  A\kern-.1667em\lower.5ex\hbox{M}\kern-.125emS}}

\def\sign(#1){(\!-\!1)^{#1}}
\def\binom(#1,#2){ (\!\!
	 \begin{array}{c} #1 \\ #2 \end{array}\!\! ) }

\title{The FORM project}

\author{J.A.M. Vermaseren\address[MCSD]{Nikhef, \\ 
        Kruislaan 409, 1098 SJ Amsterdam, The Netherlands}%
        \thanks{Currently on leave at Institut f\"ur Theoretische 
		Teilchenphysik, Universit\"at Karlsruhe D-76131 Karlsruhe, Germany}}
       
\runtitle{The FORM project}
\runauthor{J.A.M. Vermaseren}

\begin{document}

\begin{abstract}
The necessity of the FORM project is discussed. Then the evolutionary needs 
in particle physics are considered, looking at the trends over the years. A 
guess is made at what will be needed in the (near) future. The whole is 
concluded with some critical remarks concerning the publication of results 
and programs.
\vspace{1pc}
\end{abstract}

\maketitle
\section{Why FORM?}

In particle theory we have categories of calculations that are particularly 
demanding on hardware and software facilities. So much so that particle 
theory has stood at the cradle of symbolic computation and also afterwards 
has made large contributions to it. Yet, as soon as a system becomes bigger 
and bigger and commercially interesting, it often leaves its origins and it 
becomes more and more difficult to influence its development.

Hence we need one or more systems of which we can influence the 
development. This way it can be optimized or close to optimized for our 
needs. The best case is if the author(s) is/are involved in our type of 
calculations. In the next best case we should be in the position to adapt a 
system by ourselves in order to avoid a very lengthy cycle of interaction 
with the authors. This asks for an open source system that is properly 
documented to make it as easy as possible for people to make additions. 
Moreover, a system should be readily available to all researchers.

FORM~\cite{FORM,ParFORM,tform} is supposed to fit these requirements. To 
some extent it has already been like this, because I have been involved in 
many types of projects that other people in particle phenomenology are also 
engaged in. Like GiNaC~\cite{GiNaC} it is an 'in house' particle theory 
project with applications to other fields of science. The fact that FORM 
isn't open source is being worked at.

One should also realize that FORM is heavily optimized for speed and the 
handling of very large expressions. Commercial systems have usually a 
different optimization target. An overwhelming fraction of commercial 
users doesn't have problems that explore the limits of what is possible. 

\section{Trends in Loops and Legs}

If one looks at the history of calculations in particle theory one sees a 
development over the years.

At first the symbolic manipulation was to combine tensors and four vector 
dotproducts and manipulate gamma matrices. This was what 
Schoonschip~\cite{Schoonschip} was designed for and also one of the first 
things that FORM could do. Because there are still people who think in 
these terms (nowadays mainly people who are not in particle phenomenology) 
FORM has been stigmatized as a program that is only suitable for particle 
physics.

An example of a reaction that was topline research in 1976~\cite{tautau}:
$\gamma p \rightarrow \tau^- \tau^+ X \rightarrow e^- \mu^+ 
\nu\overline{\nu}\nu\overline{\nu} X$.

{\footnotesize
\begin{verbatim}
 *
 *   gamma+proton -> tau- tau+ X -> e- (nu nubar)
 *                               mu+ (nu nubar) X
 *   Narrow width approximation and full 
 *   spin-spin correllations.
 *
 S   mtau,mmu,me,mnut,mnum,mnue;
 I   j1,j2,j3,j4,e1;
 V   pa,pb,q1,q2,e2,p1,p2,p3,p4,p5,p6,pe,pm;
 
 L   F = 
   (g_(1,p1)+mnut*gi_(1))*
   (g_(2,pe)+me*gi_(2))*
   (g_(3,p4)+mnum*gi_(3))*(
     +g_(1,j1)*g7_(1)*(g_(1,q1)+mtau*gi_(1))
       *g_(1,e1) *(g_(1,q1)-g_(1,pa)
          +mtau*gi_(1))*g_(1,e2)
       *(-g_(1,q2)+mtau*gi_(1))
       *g_(1,j2)*g7_(1)*(-1/2)/q1.pa
       *g_(2,j1)*g7_(2)*g_(3,j2)*g7_(3)
     +g_(1,j1)*g7_(1)*(g_(1,q1)+mtau*gi_(1))
       *g_(1,e2)*(g_(1,pa)-g_(1,q2)
          +mtau*gi_(1))*g_(1,e1)
       *(-g_(1,q2)+mtau*gi_(1))
       *g_(1,j2)*g7_(1)*(-1/2)/q2.pa
       *g_(2,j1)*g7_(2)*g_(3,j2)*g7_(3)
   )*
   (g_(1,p6)-mnut*gi_(1))*
   (g_(2,p3)-mnue*gi_(2))*
   (g_(3,pm)-mmu*gi_(3))*(
     +g_(1,j4)*g7_(1)*(-g_(1,q2)+mtau*gi_(1))
       *g_(1,e2)*(g_(1,q1)-g_(1,pa)
          +mtau*gi_(1))*g_(1,e1)
       *(g_(1,q1)+mtau*gi_(1))
       *g_(1,j3)*g7_(1)*(-1/2)/q1.pa
       *g_(2,j3)*g7_(2)*g_(3,j4)*g7_(3)
     +g_(1,j4)*g7_(1)*(-g_(1,q2)+mtau*gi_(1))
       *g_(1,e1)*(g_(1,pa)-g_(1,q2)
          +mtau*gi_(1))*g_(1,e2)
       *(g_(1,q1)+mtau*gi_(1))
       *g_(1,j3)*g7_(1)*(-1/2)/q2.pa
       *g_(2,j3)*g7_(2)*g_(3,j4)*g7_(3)
   )/2^16;
 Trace4,3;
 Trace4,2;
 Trace4,1;
 id  q1.q1 = mtau^2;
 id  q2.q2 = mtau^2;
 id  pa.pa = 0;
 Print +f +s;
 .end

Time =    0.21 sec    Generated terms =    1992
            F         Terms in output =     176
                      Bytes used      =    8552
\end{verbatim}
}

Next came the manipulation of loop integrals. At first the one loop 
integrals and their reduction to scalar loop integrals. Here is a very 
advanced example of a Feynman diagram calculated around the year 
1983~\cite{fivepoint}:

\begin{center}
\begin{picture}(120,66)(0,0)
\SetScale{0.6}
	\ArrowLine(190,10)(10,10)
	\ArrowLine(10,100)(190,100)
	\Photon(55,10)(55,100){4}{8}
	\Photon(145,10)(145,55){4}{4}
	\Photon(145,55)(145,100){4}{4}
	\DashLine(145,55)(190,55){4}
	\Vertex(55,10){2}
	\Vertex(55,100){2}
	\Vertex(145,10){2}
	\Vertex(145,55){2}
	\Vertex(145,100){2}
\end{picture}
\end{center}

Then (1989-2000)~\cite{eerus,mincer1,mincer2,matad,moments} came also the 
manipulation of the three loop propagator graphs, done by many relations 
based on integration by parts~\cite{ibp}. The expansions in terms of 
$\epsilon$ posed additional requirements. These were the days of version 2 
of FORM.

In the late nineties new trends were emerging. Not only new techniques for 
the rewriting of the diagrams in terms of master integrals were 
developed~\cite{tarasov,remiddi,laporta}, but also new methods for the 
treatment of the integrals themselves saw the daylight. Most notoriously 
methods with nested sums~\cite{hsums} and harmonic 
polylogarithms~\cite{hpol}. The rewriting started to need the solving of 
large sets of equations. Additionally the occasional calculation of more 
and more complicated color factors~\cite{color} required some additional 
types of topological pattern matching. These last methods have not yet been 
explored very much as they will be needed mostly when there are very many 
loops. But they can also be very useful for complicated tensor algebra. 
Also the use of very large tables inspired new and original solutions. 
Version 3 of FORM has these methods in mind.

By now the traces of the gamma matrices form just a very small corner in 
the space of its capabilities.

Several types of calculations will always be at the limits of what is 
possible. A good example is the project at Karlsruhe of Baikov, Chetyrkin 
and K\"uhn~\cite{baikov}. Given more power, they can do deeper 
calculations. Hence it is important to have FORM as powerful as possible. 
This is adressed by the ParFORM~\cite{ParFORM} project and more recently 
this has led to TFORM~\cite{tform}.

Another example of something that is in principle open ended is the 
expression of multiple zeta values into a minimal set of 
variables~\cite{mzv}. This is 
the status at 2007\footnote{We make no attempt to be exhaustive here.}

{\footnotesize
\begin{center}
\begin{tabular}{|c|c|c|c|c|} \hline
	weight  &  variables  &    equations  &   left     &  time(sec) \\ \hline
      1     &      2      &        1      &    2       &            \\
      2     &      6      &       10      &    1       &            \\
      3     &     18      &       38      &    1       &            \\
      4     &     54      &      138      &    1       &            \\
      5     &    162      &      462      &    2       &            \\
      6     &    486      &     1486      &    2       &            \\
      7     &   1458      &     4730      &    4       &        8.8 \\
      8     &   4374      &    15110      &    5       &       46.8 \\
      9     &  13122      &    48558      &    8       &      306   \\
     10     &  39366      &   158602      &   11       &     2382   \\
     11     & 118098      &   515858      &   18       &    28906   \\
     12     & 354294      &  1669610      &   25       &  1243191   \\ \hline
\end{tabular}\vspace{2mm} \\
\small
Time is real time on a computer with 8 Xeon cores at 2.3 GHz and 2 Gbytes 
of memory per core, running TFORM.
\end{center}
}

The recent developments in massless multi-particle one loop 
amplitudes~\cite{superrel} hasn't led yet to particular symbolic projects. 
It is not clear whether it is needed. The methods with sector 
decomposition~\cite{sectordeco} are under development and again, it isn't 
100\% clear whether they need new developments in the symbolic sector. 
Possibly internal capabilities for treating combinations of sums, theta 
functions, delta functions and the splitting of factors with denominators 
could speed up the nested sums considerably. This would be very useful for 
the current methods based on the Mellin-Barnes approach~\cite{MB}. This is 
however not entirely trivial, unless it will be too specific for a single 
problem. The automatic calculations as in GRACE~\cite{GRACE} and 
CompHEP~\cite{CompHEP} can definitely use some new facilities in the field 
of code simplification.

I am probably forgetting a few things here.

\section{The current status}

What is the status of the FORM project?

First in the field of manpower. Of course I myself work at the moment 
almost full time on FORM. In addition Misha Tentyukov is involved in the 
ParFORM project. He has been making other additions in the 
past~cite{External}. At the moment Jens Vollinga is on a three year postdoc 
position at Nikhef which involves also work for FORM. He has already made 
the code for systems independent .sav files and is working on a failsafe 
system that (at some cost of course) allows one to set up checkpoints from 
which one can restart after a computer failure. He is also setting up a 
framework for documentation and provide better installation using the 'make 
install' approach.
As part of a project grant of FOM we have the money to get a programmer for 
18 months (starting in the autumn) to help with the project of code 
simplification. This is to facilitate the FORM version of 
GRACE~cite{GraceForm}.

Over the past few years TFORM~\cite{tform} has been developed as a 
complement to ParFORM~\cite{ParFORM}. Each of the two has restrictions. 
ParFORM can operate on clusters and TFORM works only on multi core systems 
with shared memory. Because of the shared memory some things are much 
easier in TFORM. Much administrative work needs only a single copy. Much 
multiple reading of files can be done with a relatively simple locking 
system. ParFORM has the advantage that clusters can have many more 
processors. But the communication is much more complicated. Optimization of 
the programs is a field of research and may need some extra manpower. The 
ideal would be a system that can use clusters of multi core machines. The 
problem is to reduce the bottlenecks so that for $N$ processors the 
execution time comes as close as possible to $1/N$ times the time needed on 
a single processor. Currently it is rather hard to get beyond $1/5$ on $8$ 
processors and beyond $1/10$ on $32$ processors. A careful study and 
inventive solutions will be needed.

Recently we have started to make use of the GMP (GNU Multiple 
Precision)~\cite{GMP} library for some of the computations with large 
integers. This is because the size of integers and fractions has become 
larger and larger and is often way beyond what was envisioned originally. 
We use only the low level routines for multiplication, division and GCD 
calculations. The gain in speed isn't impressive though, because the 
algorithms inside FORM are rather decent (especially after the improvements 
found in the end of 2006). But the GMP can do some things more efficiently 
because it has some assembler routines and in assembler one can do a number 
of things far more efficiently than in C. The need to convert from FORM 
notation to GMP notation introduces an overhead. We still need to 
experiment with what is the optimal size below which we should use the 
original routines and above which we should use the GMP library. If one 
does calculations that involve fractions with very large integers (like 
hundreds of digits) one will find that the more recent versions (2007 and 
on) of FORM are noticeably faster.

\section{What to expect and hope for}

The systems of equations that need to be solved are asking often for 
capabilities with rational polynomials. This is particularly the case with 
the Laporta algorithm~\cite{laporta}. It is something that FORM doesn't 
have currently. Hence it has rather high priority to build this in. And to 
build this in in a rather efficient way. There exist publicly available 
libraries for the manipulation of polynomials in a single variable, some of 
them claiming great efficiency, but there are no equivalent libraries for 
polynomials in many variables. In addition there is the problem of 
notation. Too much time spent on conversion will not be beneficial. 
Currently the problem is under study. Most univariate algorithms (in 
particular the GCD) have been implemented in various methods. This is by 
now reasonably fast. Factorization is less urgent, but can come in handy 
when constructing a system for simplification.

It is also important to deal with multivariate rational polynomials 
efficiently when one likes to create a system for computing Gr\"obner 
bases. There are however several ways to deal with polynomials and each way 
needs its own solution:
\begin{itemize}
\item Small polynomials: when they take a small amount 
of space they can be kept inside the argument of a function. There may be 
billions of such polynomials. They should be treated inside the regular 
workspace. Univariate polynomials will usually be in this category. An 
improvement in efficiency will be to tabulate a number of them. This is 
especially the case for factorization which is relatively expensive.
\item Intermediate polynomials: these could be 
handled by means of memory allocations as is done with the dollar 
variables. One could have hundreds or even thousands of them. Typically not 
billions.
\item Large polynomials: These are complete expressions that could have 
billions of terms. Calculating their GCD would have to use the same 
mechanisms as by which expressions are treated. There should be only very 
few of these.
\end{itemize}

An example of something that works already: PolyRatFun is an experimental 
statement that is similar to the PolyFun statement, but now the function 
needs two arguments: a numerator and a denominator.

{\small
\begin{verbatim}
    Symbols x,y;
    CFunction pacc;
    PolyRatFun pacc;
    L   F = pacc(x^2+x-3,(x+1)*(x+2))*y
           +pacc(x^2+3*x+1,(x+3)*(x+2))*y^2;
    Print +s;
    .sort

   F =
     + y*pacc(x^2 + x - 3,x^2 + 3*x + 2)
     + y^2*pacc(x^2 + 3*x + 1,x^2 + 5*x + 6)
      ;

    id  y = 1;
    Print;
    .end

   F =
      pacc(2*x^2 + 4*x - 4,x^2 + 4*x + 3);
\end{verbatim}
}

Sometimes one would like to have quick private additions for things that 
are extremely hard to program at the FORM level. Such things are often 
either of combinatoric nature or special patterns. It is of course 
impossible to forsee what some people will need. Hence FORM should be 
structured in such a way that it is possible to make such additions, even 
though this won't be for beginners. The first requirement for this is a 
good documentation of the inner workings, including a number of examples. 
The second requirement is code that can be understood and is structured 
properly. Due to these two requirements FORM hasn't been released yet as 
open source. We hope to be this far in about two years time.

As mentioned before, we like to have a way to introduce code 
simplification. This would be relevant for all outputs that would need 
further numerical evaluation in the languages Fortran and C. If it is 
possible we would like to extend this to the regular output for as far as 
factorization is concerned. Already some things can be done at the FORM 
level, but this is usually rather slow. One can for instance make a 
procedure `tryfactor' which would work like

{\small
\begin{verbatim}
    #do i = -100,100
    #call tryfactor(acc,x+`i')
    #enddo
    B acc;
    Print;
\end{verbatim}
}

and the answer might be like

{\small
\begin{verbatim}
    +acc(x-27)*acc(x+6)*acc(x+67)*(.......)
\end{verbatim}
}

Because this is very slow and requires guessing the factors, this is far 
from ideal.

Something that the community should think about: At a given moment Nikhef, 
and/or I may not be able to take the responsibility for FORM any longer. 
Which institute/individual(s) can take over this responsibility? Would FORM 
disappear? It is not a good idea to depend on the free time of some 
individuals. The open source project may help, but this is probably not 
sufficient. There should be a professional commitment. Of course, if 
someone can come with a better product, evolution will take its course. But 
that would require a large investment as well. Good ideas are needed here, 
because it doesn't look like CERN (which would be the most natural choice) 
is volunteering.

\section{Some critical remarks}

Some people prefer to use expensive commercial systems and give their 
results in terms of routines for these systems. I believe this to be very 
shortsighted.

Years ago we had a preprint system, and only the top universities would get 
the preprints and be up to date. Poor universities would not be able to be 
up to date and hence meaningful up to date research could only be done at a 
limited number of places.

Now with the internet, everybody can be up to date and meaningful research 
can be done everywhere. If however we present our results in the form of 
programs for very expensive software systems, we take a big step back. It 
is a form of elitism.

I am not pushing here for my own program. What I want to say is that it is 
in the interest of science that all results are freely available and freely 
accessible, and that the threshold for using the results is as low as 
possible.
\vspace{3mm}

If someone isn't happy with the facilities offered by the free software, 
spend some effort or resources on helping with providing such facilities. 
That is something that everybody can benefit of.

Another thing (Remember Babylon):

The situation becomes really chaotic when there are many complementary 
results from different authors set up for different systems. It becomes 
rapidly impossible to combine such results.

\section{Acknowledgements}

The author wishes to thank the Humboldt foundation for its generous support 
and the university of Karlsruhe for its kind hospitality.


\end{document}